\documentclass{llncs}

\usepackage{hyperref} % for URL formatting
\usepackage{natbib} % for citing authors automatically
\usepackage{graphicx} % for inserting images
\usepackage{amsmath} % for align environment

\newcommand{\coals}[1]{\mathcal{N}^{#1}}
\newcommand{\avg}{\textup{avg}}

\newcommand{\FR}{\textup{FR}}
\newcommand{\FO}{\textup{FO}}
\newcommand{\EO}{\textup{EO}}
\newcommand{\SF}{\textup{SF}}
\newcommand{\EQ}{\textup{EQ}}
\newcommand{\AL}{\textup{AL}}

\author{Luke Harold Miles}
\title{A Simulator for Hedonic Games}
\institute{University of Kentucky}

\begin{document}

\maketitle

\section{A Story}

Mr Holt, the kindergarten teacher, gives his class these instructions:
\begin{quote}
  Hello class, The Metropolitan Museum of Art has a sudden shortage of
  sculptures and needs several new ones to fill its shelves. Please break into
  groups so that each group can build a Lego tower. The director of the museum
  will be here in an hour to pick up the towers and put them in the museum with
  your names on them. Please do the best job you can; you don't want to be
  professionally embarrassed.
\end{quote}
Each kindergartener wants to be in a group with her friends, but she also wants
her friends to be happy in the group; she doesn't want her friends to be
miserable.
The graph below is a map of who is friends with whom in the small class. Notice
that $a$ would have more friends in the group $\{a, b, c, d, e\}$ than
$\{a, b, c, d\}$, but maybe $a$ doesn't want $e$ to be in the group because $a$
knows that would make $b$, $c$, and $d$ less happy. Strangely, $a$ prefers
$\{a, b, c, d\}$ to $\{a, b, c, d, e\}$.

You can imagine that the kindergarteners might try to choose the best group in
some other way. The class would split into groups one way, but then people
would be unhappy and keep changing their groups. How can we model all this? How
could we easily visualize all this?

\centerline{\includegraphics[width=2in]{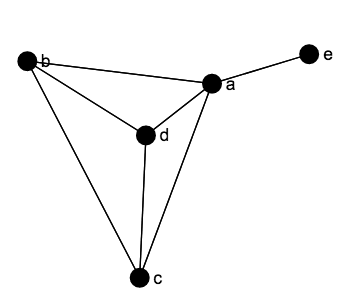}}

\section{Hedonic Games}

Below is the original definition of a hedonic game. Hedonic games
\citep*{banerjee2001core} were invented to model the formation and reformation
of groups.
\begin{definition}
  \citep*{banerjee2001core} A \textbf{coalition formation game} is a pair
  $G = (N, (\succeq_i)_{i \in N})$, where $N$ is a finite set of players and
  for every $i \in N$, $\succeq_i$ is a reflexive, complete, and transitive
  binary relation on $\coals{i} = \{C \in 2^N : i \in C\}$.
  If $C,D \in \coals{i}$ and $C \succeq_i D$ and $D \not\succeq_i C$, then we write $C \succ_i D$.
\end{definition}
\begin{definition}
  \citep*{banerjee2001core} A \textbf{coalition structure}
  $\Gamma = \{C_1, \dots, C_k\}$ is a partition of $N$.
  The coalition containing a player $i \in N$ is denoted $\Gamma(i)$.
  Any subset of $N$ is called a coalition.
\end{definition}
That's a very minimal definition, and these most general hedonic games don't
have many computationally useful properties. For that reason, several subclasses
of hedonic games have been invented and studied. First though, let's look at
stability.

\subsection{The Core}

If Mr Holt were assigning groups, instead of letting the kids form their own
groups, then he might want a way to predict if a given partition will stick
before he actually moves people around. ``Will the students stay in their groups
or will they form new ones?'' There are many ways you can ask the question ``Is
this coalition formation stable?'' Seven good ways are mentioned in
\citep*{nguyen2016altruistic}. One of the most important ways to ask the
question (and the focus of the survey \citep*{woeginger2013core}) is ``Is this
this coalition formation core stable?''.

\begin{definition}
In a hedonic game $G$ with a partition $\Gamma$, if there is a nonempty set
$C \subseteq N$ where $\forall i \in C: C \succ_i \Gamma(i)$, then we say that
$C$ blocks $\Gamma$, or $C$ is a \textbf{blocking coalition} in $\Gamma$.
If $\Gamma$ cannot be blocked, then it is called \textbf{core stable}. The set
of core stable partitions for a game $G$ is called the \textbf{core} of $G$.
\end{definition}

\section{Varieties of Hedonic Games}

In the below paragraphs, $n = |N|$ is the number of players, $i$ is a player in
$N$, and $C,D \in \coals{i}$ are coalitions which contain $i$.

\subsection{Fractional Hedonic Games}

\citep*{aziz2014fractional} In \textbf{fractional hedonic games}, $i$ assigns some real value $v_i(j)$ to every
player $j \in N$. It's assumed that $v_i(i) = 0$.\footnote{
  Raising your own score is equivalent to lowering everyone else's score.
  Lowering your own score is equivalent to raising everyone else's score.}
We say $C \succeq^{\FR}_i D$ if $u^{\FR}_i(C) \geq u^{\FR}_i(D)$, where
$$u_i^{\FR}(C) = \sum_{j \in C} v_i(j). $$

A fractional hedonic game is called \textbf{simple} if
$\forall i,j \in N: v_i(j) \in \{0,1\}$
and is called \textbf{symmetric} if
$\forall i,j \in N: v_i(j) = v_j(i)$.
\citeauthor*{aziz2014fractional} show that even in fractional hedonic games which are both simple 
and symmetric, the core is sometimes
empty and that checking core emptiness is $\Sigma_2^p$-complete.

\subsection{Friend and Enemy Oriented Hedonic Games}

\citep*{dimitrov2006simple} In both of these kinds of games, $i$ splits the
other players in $N$ into a set of friends, $F_i$, and a set of enemies, $E_i$.

In \textbf{friend-oriented games}, $i$ prefers coalitions with more friends and breaks
ties by considering the number of enemies. In other words,
\begin{align*}
       & C \succeq^{\FO}_i D \\
  \iff & |C \cap F_i| > |D \cap F_i| ~\lor~ \left( |C \cap F_i| = |D \cap F_i| ~\land~ |C \cap E_i| \leq |D \cap E_i| \right) \\
  \iff & u_i^{\FO}(C) \geq u_i^{\FO}(D), \\
  \textup{where } & u_i^{\FO}(C) = n|C \cap F_i| - |C \cap E_i|.
\end{align*}
So if $C$ has 8 of $i$'s friends and 600 of $i$'s enemies and $D$ has 7 of
$i$'s friends and 0 of $i$'s enemies, then $i$ would still rather be in $C$.

In \textbf{enemy-oriented games}, $i$ tries to minimize enemies and only considers
friends to break a tie. In other words,
\begin{align*}
       & C \succeq^{\EO}_i D \\
  \iff & |C \cap E_i| < |D \cap E_i| ~\lor~ \left( |C \cap E_i| = |D \cap E_i| ~\land~ |C \cap F_i| \geq |D \cap F_i| \right) \\
  \iff & u_i^{\EO}(C) \geq u_i^{\EO}(D), \\
  \textup{where } & u_i^{\EO}(C) = |C \cap F_i| - n |C \cap E_i|.
\end{align*}

\citeauthor*{dimitrov2006simple} show that the core is guaranteed to be
non-empty in both kinds of games. However, finding a core stable partition is
NP-hard in enemy-oriented games\footnote{
  More precisely, if you could always find a core stable coalition structure in
  polynomial time, then you could also find the largest clique in any
  (undirected, unweighted) graph in polynomial time.}
but polynomial time in friend-oriented games.

\subsection{Altruistic Hedonic Games}

\citep*{nguyen2016altruistic} As in friend and enemy oriented hedonic games,
$i$ divides the other players into friends, $F_i$, and enemies, $E_i$. The idea
is that a player wouldn't want to be in a coalition $C$ where his friends were
miserable, even if $C$ had all of his friends and none of his enemies.

Three levels of altruism are considered. Let $\avg(S) = \sum_{x \in S} x / |S|$
denote the average of a multiset of numbers. And, as above, the utilities $u_i$
are defined so that $C \succeq_i D \iff u_i(C) \geq u_i(D)$.

In \textbf{selfish-first altruistic games}, a player cares most about his own happiness
and uses his friends' preferences to break ties. `Happiness' here means the
friend-oriented score.  This is distinct from friend-oriented games in that a
tightly connected coalition $C$ with 6 friends and 3 enemies is preferred to a
sparse coalition $D$ with 6 friends and 3 enemies, because $i$'s friends in $C$
are happier than $i$'s friends in $D$.
$$u_i^{\SF}(C) = n^5 u_i^{\FO}(C) + \avg({u_j^{\FO}(C) : j \in C \cap F_i}).$$
In \textbf{equal-treatment altruistic games}, a player takes his and all his friends'
opinions into account equally when evaluating a partition:
$$u_i^{\EQ}(C) = \avg(u_j^{\FO}(C) : j \in C \cap F_i \cup \{\i\}).$$
And in \textbf{altruistic-treatment altruistic games} (i.e., truly altruistic games), a
player prefers coalitions where his friends are happy and breaks ties by
considering his own happiness.
$$u_i^{\AL}(C) = u_i^{\FO}(C) + n^5 \avg({u_j^{\FO}(C) : j \in C \cap F_i}).$$

\citeauthor*{nguyen2016altruistic} show that selfish-first altruistic games
always have an nonempty core. Whether equal-treatment altruistic games and
truly altruistic games ever have empty cores are open questions. I suspect
that the core is always nonempty in both games.

\section{The Simulator}

I wrote software to simulate hedonic games and put in on the internet.
You can draw graphs, choose partitions, choose several different player types,
and check the stability of the partition under several different measures.
Hopefully this will help others and myself quickly understand different
hedonic games and speed up the process of finding stable partitions.

\newpage
\begin{center}
  \url{http://lukemiles.org/hedonic-games} \\
  \includegraphics[width=3in]{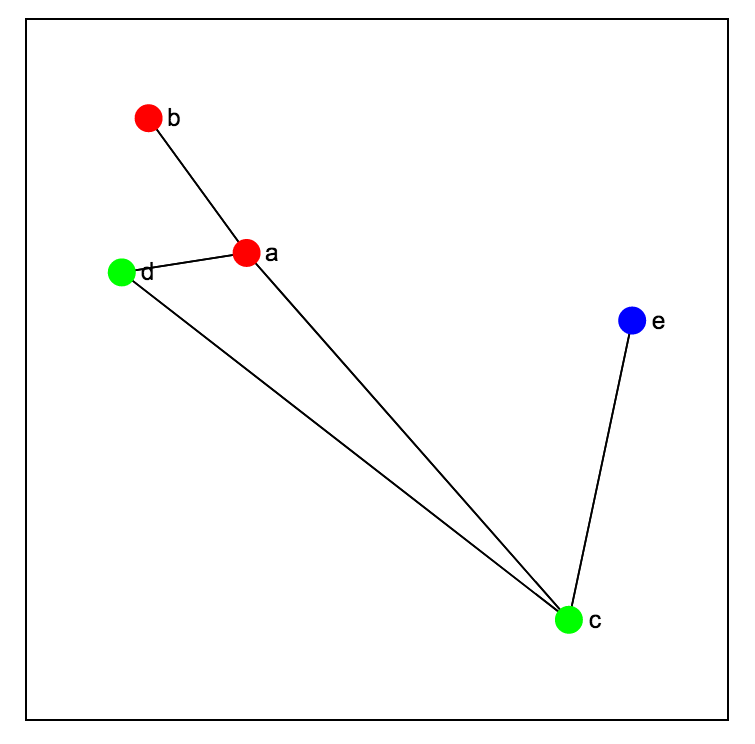} \\
  \includegraphics[width=3in]{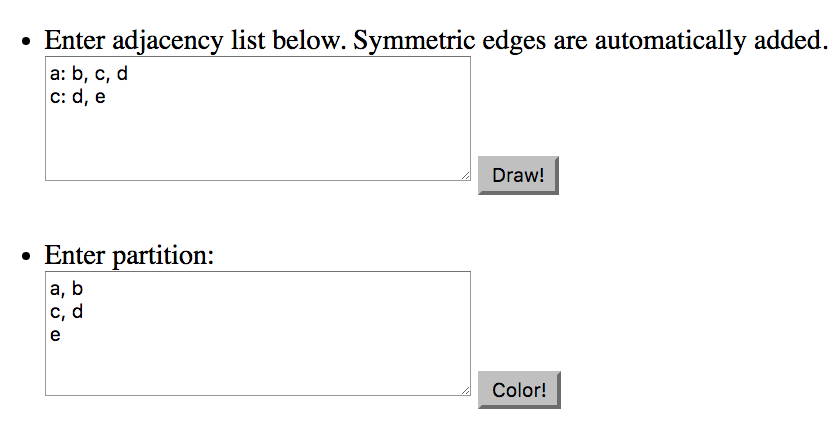} \\
  \includegraphics[width=3in]{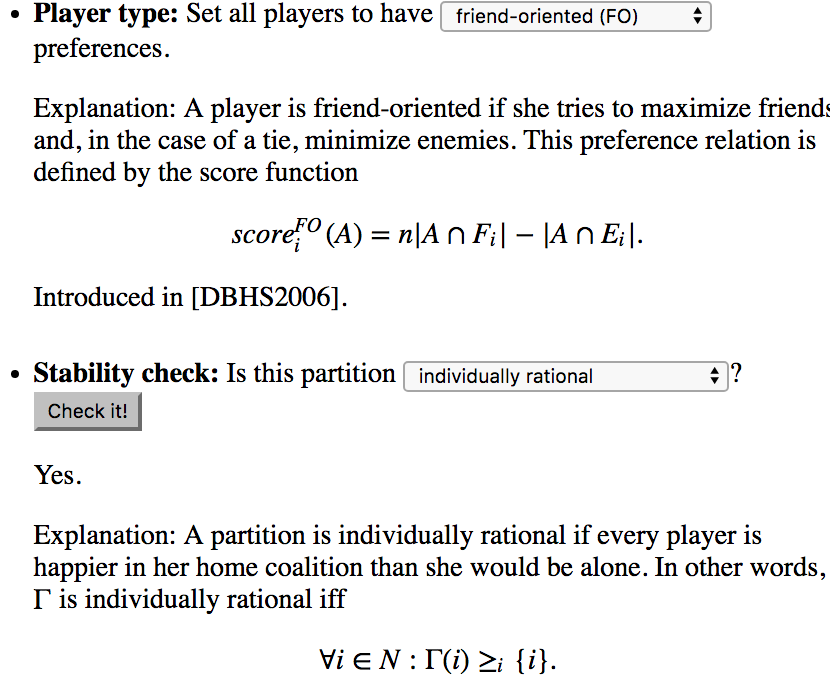} \\
  The website works better on laptops than smartphones. Updates may have been
  made to the website since this arXiv version was uploaded.
\end{center}

\bibliographystyle{plainnat}
\bibliography{hedonic-games}

\begin{thebibliography}{5}
\providecommand{\natexlab}[1]{#1}
\providecommand{\url}[1]{\texttt{#1}}
\expandafter\ifx\csname urlstyle\endcsname\relax
  \providecommand{\doi}[1]{doi: #1}\else
  \providecommand{\doi}{doi: \begingroup \urlstyle{rm}\Url}\fi

\bibitem[Aziz et~al.(2014)Aziz, Brandt, and Harrenstein]{aziz2014fractional}
Haris Aziz, Felix Brandt, and Paul Harrenstein.
\newblock Fractional hedonic games.
\newblock In \emph{Proceedings of the 2014 international conference on
  Autonomous Agents \& Multi-Agent Systems (AAMAS)}, 2014.

\bibitem[Banerjee et~al.(2001)Banerjee, Konishi, and
  S{\"o}nmez]{banerjee2001core}
Suryapratim Banerjee, Hideo Konishi, and Tayfun S{\"o}nmez.
\newblock Core in a simple coalition formation game.
\newblock \emph{Social Choice and Welfare}, 2001.

\bibitem[Dimitrov et~al.(2006)Dimitrov, Borm, Hendrickx, and
  Sung]{dimitrov2006simple}
Dinko Dimitrov, Peter Borm, Ruud Hendrickx, and Shao~Chin Sung.
\newblock Simple priorities and core stability in hedonic games.
\newblock \emph{Social Choice and Welfare}, 2006.

\bibitem[Nguyen et~al.(2016)Nguyen, Rey, Rey, Rothe, and
  Schend]{nguyen2016altruistic}
Nhan-Tam Nguyen, Anja Rey, Lisa Rey, J{\"o}rg Rothe, and Lena Schend.
\newblock Altruistic hedonic games.
\newblock In \emph{Proceedings of the 2016 international conference on
  Autonomous Agents \& Multi-Agent Systems (AAMAS)}, 2016.

\bibitem[Woeginger(2013)]{woeginger2013core}
Gerhard~J Woeginger.
\newblock Core stability in hedonic coalition formation.
\newblock In \emph{Proceedings of SOFSEM 2013: Theory and Practice of Computer
  Science}, 2013.

\end{thebibliography}

\end{document}